\documentclass[runningheads]{llncs}

\usepackage{graphicx}

\usepackage{biblatex}
\usepackage{hyperref}  
\usepackage{url} 
\usepackage{amsfonts} 
\usepackage{nicefrac}

\usepackage{microtype} 
\usepackage{comment}
\usepackage{caption}
\usepackage{color}
\usepackage{subcaption}
\usepackage{multirow}
\usepackage{booktabs}
\usepackage{pifont}%
\usepackage{float}
\usepackage{amsmath}

\begin{document}
\title{TRAKO: Efficient Transmission of Tractography Data for Visualization}
\author{Daniel Haehn\inst{1}\and
Loraine Franke\inst{1}\and Fan Zhang\inst{2}\and Suheyla Cetin Karayumak\inst{2} \\ \and Steve Pieper\inst{3} \and Lauren O'Donnell\inst{2} \and Yogesh Rathi\inst{2}}
\authorrunning{D. Haehn et al.}
\institute{University of Massachusetts Boston \and 
Harvard Medical School \and Isomics, Inc.}
\maketitle              %
\begin{abstract}
Fiber tracking produces large tractography datasets that are tens of gigabytes in size consisting of millions of streamlines. Such vast amounts of data require formats that allow for efficient storage, transfer, and visualization. We present TRAKO, a new data format based on the Graphics Layer Transmission Format (glTF) that enables immediate graphical and hardware-accelerated processing. We integrate a state-of-the-art compression technique for vertices, streamlines, and attached scalar and property data. We then compare TRAKO to existing tractography storage methods and provide a detailed evaluation on eight datasets. TRAKO can achieve data reductions of over 28x without loss of statistical significance when used to replicate analysis from previously published studies.

\end{abstract}

\section{Introduction}

\begin{figure}[t!]
    \centering
    \includegraphics[width=.45\textwidth]{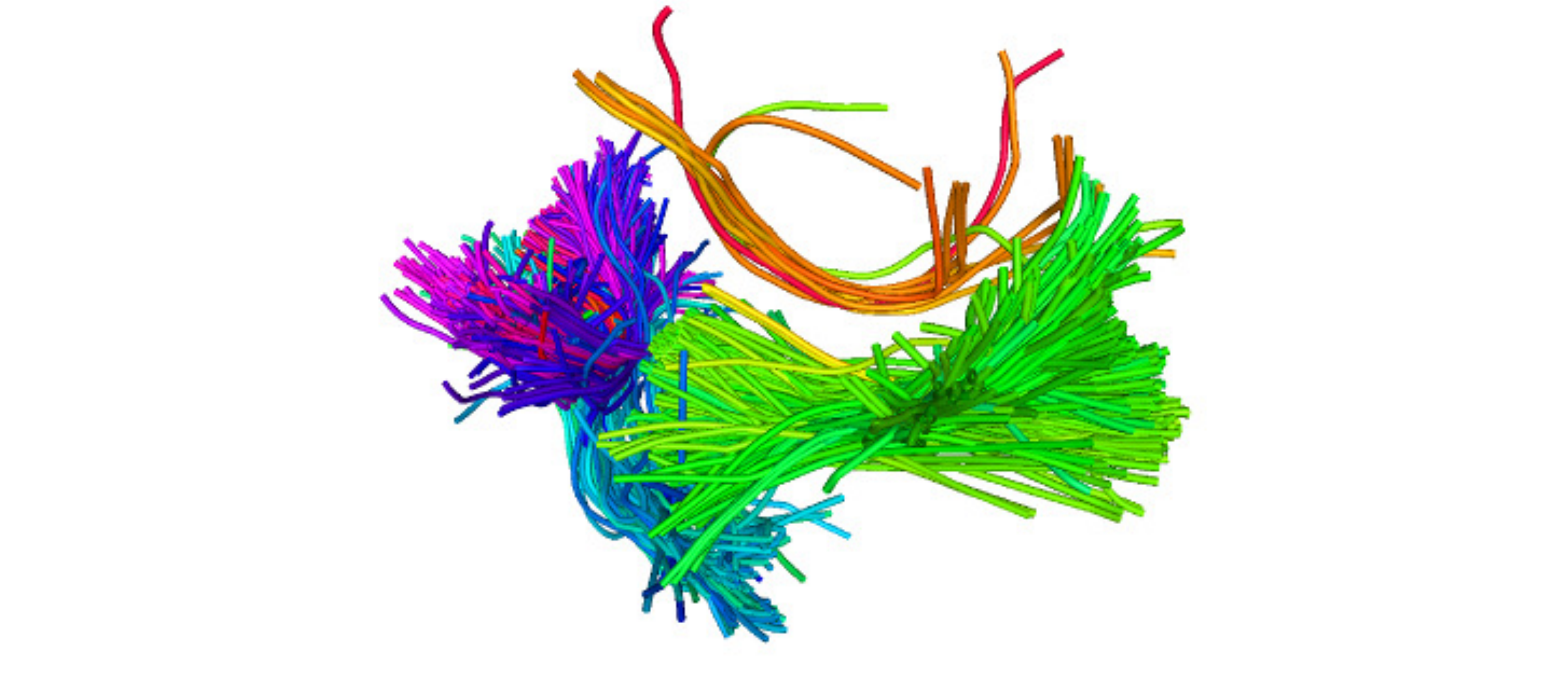}
    \includegraphics[width=.45\textwidth]{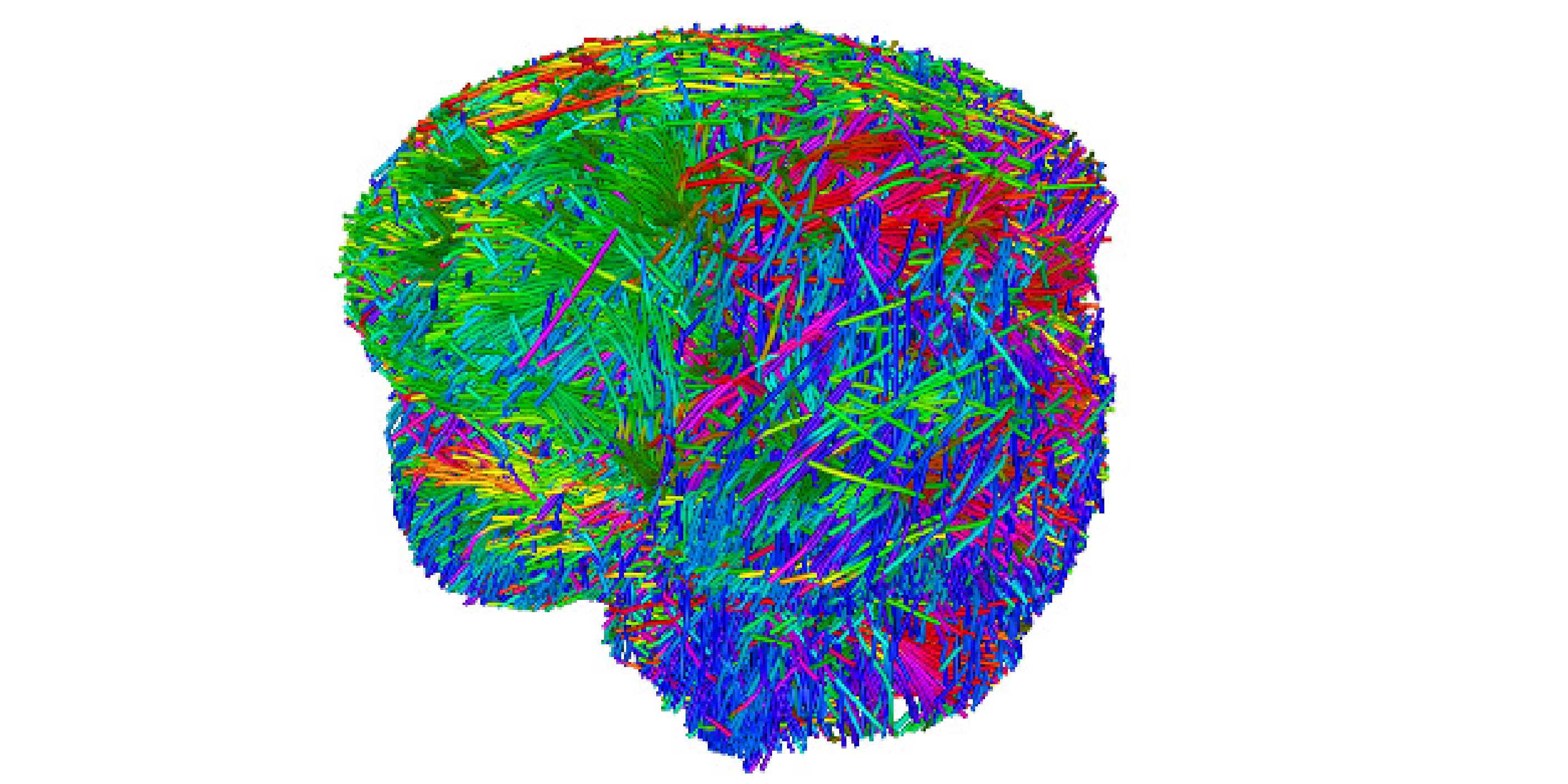}
    \caption{Examples of diffusion tractography fiber tracts. (left) separate fiber clusters, (right) wholebrain tractography. Individual tracts are colored by anatomical orientation.}
    \label{fig:examples}
\end{figure}

Diffusion-weighted magnetic resonance imaging (MRI) allows estimation of the brain's white matter properties~\cite{basser1994mr}. Fiber tracking methods~\cite{basser2000vivo} then produce clusters of streamlines corresponding to 3D fiber bundles (Fig.~\ref{fig:examples}). Each fiber in these bundles is a line with a collection of $x,y,z$ coordinates, typically represented using 32-bit floating point numbers. Researchers may attach scalars to these coordinates (per-vertex) to record values such as estimates of local tissue integrity. These values can be of arbitrary dimension, size, and data type. Researchers may also attach many different property values to individual streamlines (per-fiber). Modern tractography studies with scalars and properties can result in datasets that are tens of gigabytes in size per subject~\cite{rheault2017visualization}. Storing such data can be expensive while transferring and processing the data for visualization can be inefficient. To optimize the costs and minimize overall delays, we need to explore compression techniques and their effect on tractography based neuroanalysis.

Currently existing compression methods are using two approaches by either reducing the number of fiber tracts in a dataset by downsampling ~\cite{gori2016parsimonious, alexandroni2017fiber, garyfallidis2012quickbundles, guevara2011robust, mercier2018progressive, liu2012unsupervised, siless2018anatomicuts, olivetti2017comparison, zhang2018anatomically} or compressing the data of individual fibers~\cite{lindstrom2014fixed, presseau2015new, chung2009efficient, kumar2016sparse, moreno2016sparse, caiafa2017multidimensional}. However, none of the existing methods approaches the problem from the perspective of optimizing storage for graphical processing, nor do they leverage recent developments in data representation and compression standards for spatial computing.
In this paper, we present TRAKO, a new tractography data format for efficient transmission and visualization. TRAKO is based on the fully extendable glTF~\cite{robinet2014gltf} container, which among other things is designed to minimize runtime processing when uploading data to a graphical processing unit (GPU). Furthermore, TRAKO applies state-of-the-art 3D geometry compression techniques which allow to explicitly control the data reduction (lossiness). In addition, TRAKO compresses vertices of each fiber tract and attached scalars and properties, an advantage over existing tractography compression methods.

We compare TRAKO against two compression schemes that are specifically designed for fiber tracts: \textit{zfib}~\cite{presseau2015new} and \textit{qfib}~\cite{mercier2020qfib}. Zfib, which is now part of the Dipy~\cite{garyfallidis2014dipy} library, reduces the number of vertices in each fiber tract but does not change the vertices itself (downsampling). Qfib is a recently presented algorithm that compresses individual vertices and allows to choose between a 8 bit and 16 bit precision. Neither zfib nor qfib support the compression of attached per-vertex scalars or per-fiber properties. In contrast, TRAKO encodes vertices and all attached values with the Draco algorithm~\cite{brettle2017introducing} that combines quantization, prediction schemes, and attribute encoding.\footnote{https://github.com/google/draco}

Most tractography compression schemes are configurable to trade-off information loss and data size. Therefore, we explore different settings of TRAKO to encode data points with the goal of sufficiently preserving accuracy for quantitative analysis. We test and evaluate the methods TRAKO, zfib, and qfib on multiple datasets to measure the loss of vertices, scalars, and properties after encoding. TRAKO reduces data sizes by a factor of 10-28x with an average error that is lower than the voxel size of the original diffusion MRI. We further perform a sensitivity analysis and replicate two previously published tractography studies with compressed versions of the original data. We find that compressed fiber tracts are very suitable for real-world processing. Finally, we publicly release all our data, code, experiments, and results\footnote{https://pypi.org/project/trako/}.

\section{Data Format}
\subsection{Structure}

The TRAKO data format with file extension .tko, is built off the Graphics Library Transmission Format (glTF)~\cite{robinet2014gltf}, a JSON-based royalty-free format for efficient transmission and loading of 3D scenes (i.e. to be the "JPEG of 3D"). glTF containers include mechanisms to store computer graphics scenes but the specification is fully extendable and flexible. 

\begin{figure}[ht!]
    \centering
    \includegraphics[width=.7\textwidth]{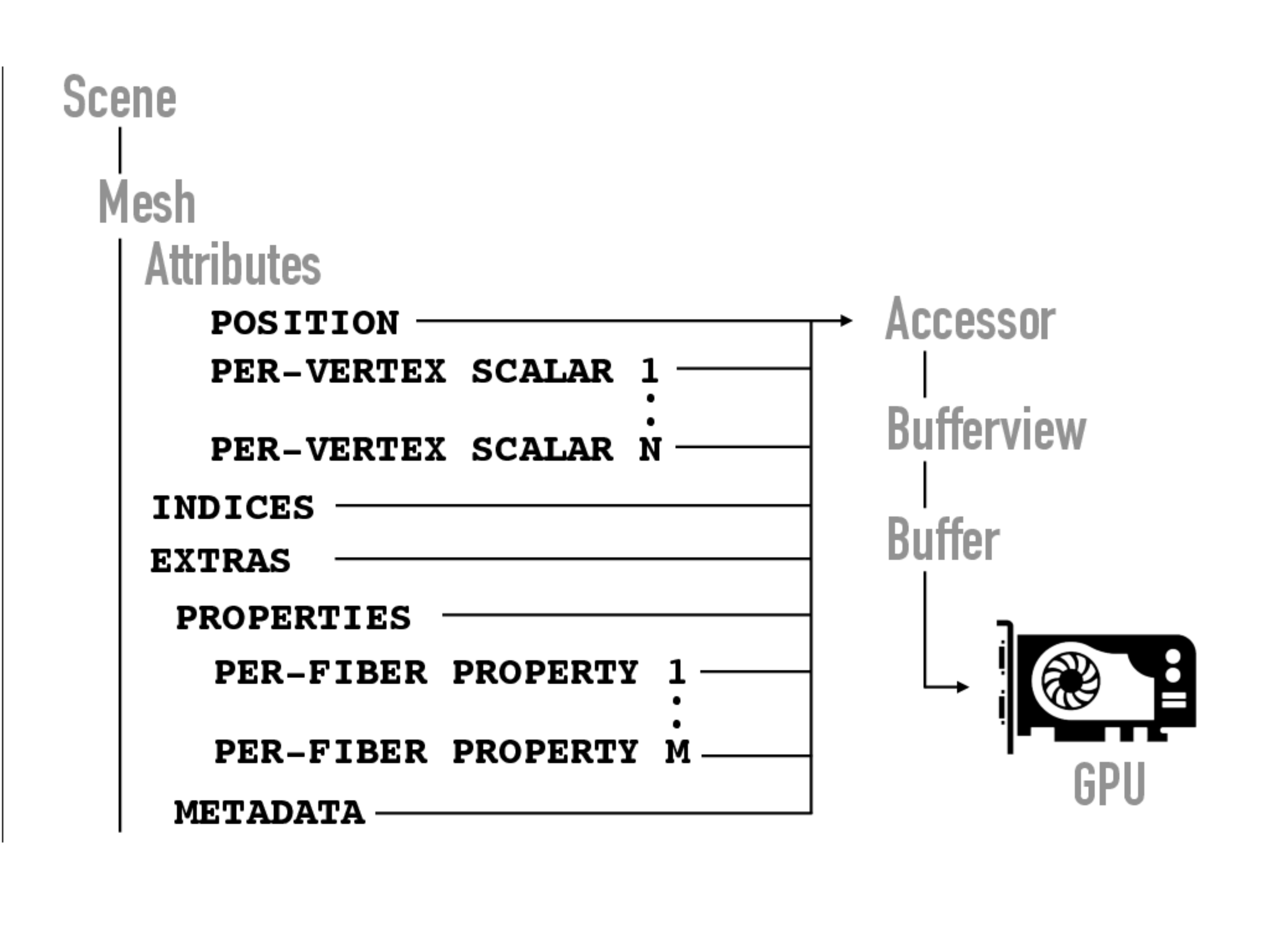}
    \caption{The TRAKO data format stores fiber tracts in a standardized glTF~\cite{robinet2014gltf} container. This way, we can use existing mechanisms such as position attributes and indices to store the streamlines as buffers. These buffers are accessible and configurable through accessors and bufferviews and are immediately ready for transmission to the GPU. glTF containers are fully extendable and allow TRAKO to support the storage of per-vertex scalars, per-fiber properties, and metadata in any format.}
    \label{fig:format}
\end{figure}

For TRAKO, we define a set of fiber tracts using the glTF mesh data structure (Fig.~\ref{fig:format}). This structure is defined with arrays of primitives corresponding directly to data required for draw calls of a GPU. Specifically, we use the \texttt{POSITION} attributes (Vec3 floats) to store the vertices of the fiber tracts and then map them to individual streamlines using the \texttt{INDICES} property. Since TRAKO files are valid glTF files as well, we can leverage the whole glTF ecosystem that includes validators, viewers, optimizers, and converters. For examples, we can convert ASCII JSON .tko-files to binary versions with existing converter tools such as the Cesium glTF Pipeline\footnote{https://github.com/CesiumGS/gltf-pipeline} or gltf-pack\footnote{https://github.com/zeux/meshoptimizer}.

\subsection{Compression}

\begin{figure}[t]
    \centering
    \includegraphics[width=.75\textwidth]{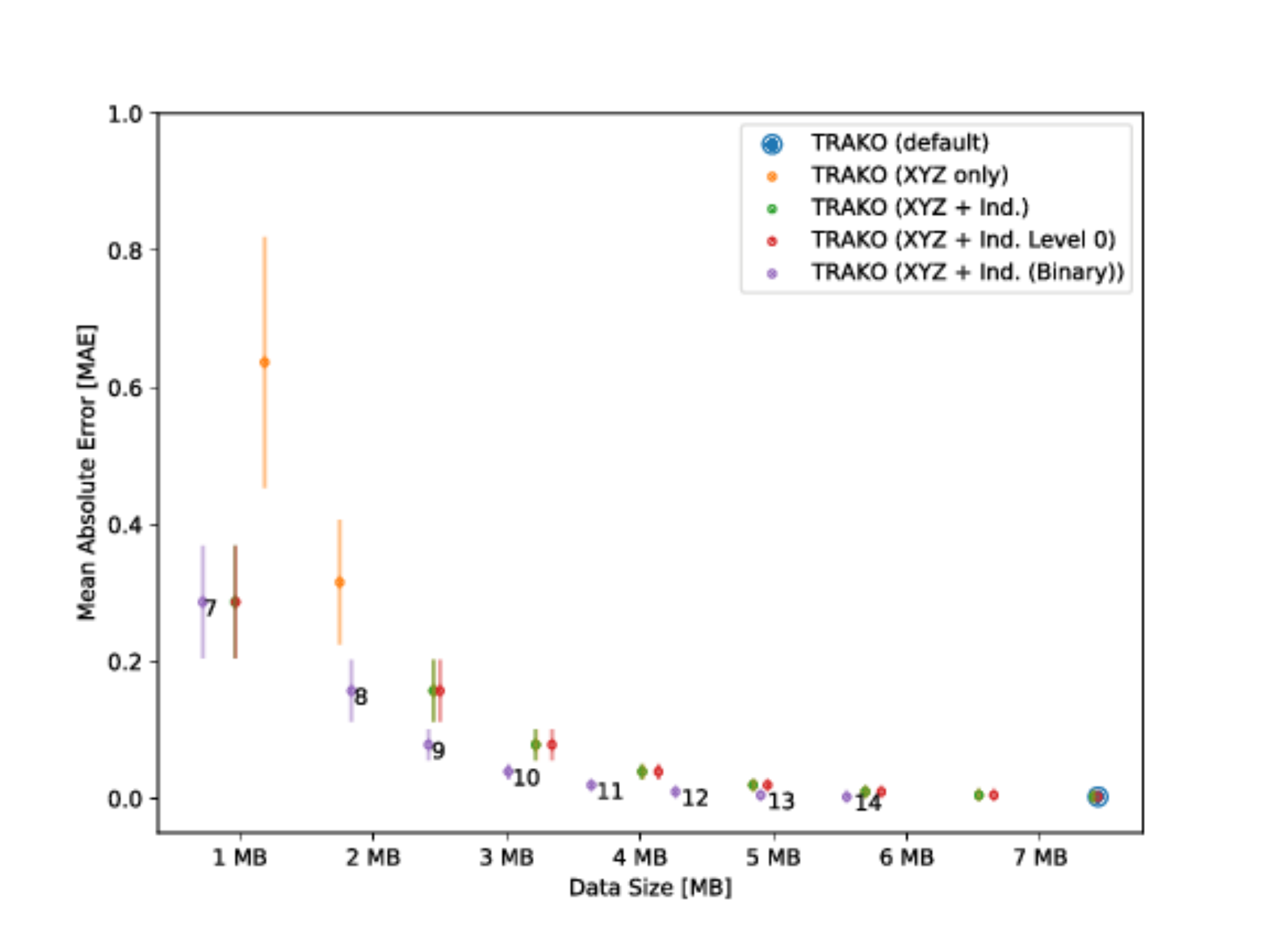}
    \caption{Parameter exploration of TRAKO on the ISMRM 2015 dataset with an original size of 34.1 Megabytes. We test the default parameters of TRAKO (blue, quantization bits (q\_bit) 14, compression level (cl) 1), a variation that only compresses the vertices (XYZ, orange), one that compresses XYZ and indices (Ind., green), the same but with compression level 0 for faster speed (red), and finally, TRAKO converted to binary using the glTF Pipeline (purple). The lower left corner indicated low errors and high compression rates. The numbers in the plot indicate the quantization bits.}
    \label{fig:ismrmbits}
\end{figure}

Internally, TRAKO leverages the Draco compression scheme that enables the compression of meshes and point cloud data by combining multiple techniques. For meshes, Draco uses the Edgebreaker algorithm~\cite{rossignac1999edgebreaker}. For point clouds, Draco offers a $k$d-tree based encoding that re-arranges all points, or a sequential encoding that preserves their order. Preserving the order is important for tractography data since we need to keep track of all vertices and any mapped values along the streamlines. We integrated Draco's sequential encoding method to TRAKO. This method combines entropy reduction using a configurable quantization rate of 1-31 bits with prediction schemes that compute differences between stored values (similar to delta encoding)~\cite{deering1999compression}.

There are two main parameters to control the compression. The quantization rate controls how many bits are used to encode individual values (default: 14). Higher rates allow for greater data precision but yield larger data sizes. We explored quantization rates in the range of 7-14 bits as part of an initial parameter exploration (Fig.~\ref{fig:ismrmbits}). The second main parameter of Draco is the compression level from 0-10. This level can be used to trade off encoding speed with better compression. Since speed is not of primary importance, we always select the maximum compression level of 10.

\subsection{File Formats}

Our TRAKO implementation supports conversion and on-the-fly compression of data (\textit{trakofy}), decompression of data (\textit{untrakofy}), and comparison of an uncompressed file to the original source file (\textit{tkompare}). These tools support various widely used tractography data formats including VTK, VTP\footnote{https://vtk.org/wp-content/uploads/2015/04/file-formats.pdf}, TCK\footnote{https://mrtrix.readthedocs.io/en/latest/getting\_started/image\_data.html}, and TRK\footnote{http://www.trackvis.org/dtk/?subsect=format} files.  In addition, we provide a reusable Python package to allow integration of TRAKO with other software systems or for extension to support other file formats.  The glTF standard itself provides a standard mechanism for embedding domain-specific data within glTF JSON structures, and there exists a wide range of extensions to support features such as advanced graphical rendering, animation, and multiple levels-of-detail\footnote{https://github.com/KhronosGroup/glTF/blob/master/extensions/README.md}. The same approach can be used with TRAKO to embed custom experimental metadata without breaking compatibility with the core standard.

\section{Evaluation and Results}

\subsection{Performance}
\begin{table}[!ht]
    \centering
    \caption{We evaluate TRAKO on eight different datasets. The top five datasets only contain streamlines and vertices (TCK format). The bottom three datasets include attached per-vertex scalars and per-fiber properties, resulting in large data sizes (VTK and VTP formats). Abbreviations: UKF - unscented Kalman Filter tractography; iFOD1: 1st order integration over fiber orientation distributions tractography; HCP - Human Connectome Project (one example young healthy adult); dHCP - Developing Human Connectome Project (one example neonate); ADHD - Attention deficit hyperactivity disorder dataset (including 30 ADHD patients and 29 healthy control subjects) }
    \label{tab:datasets}
    \resizebox{\linewidth}{!}{
    \begin{tabular}{lrrrcccr}
        \toprule
        \textbf{Dataset} & \textbf{Streamlines} & \textbf{Vertices} & \textbf{Tracking} &\textbf{Scalars} & \textbf{Properties}& \textbf{Format} & \textbf{Size} \\
        \midrule
        qfib-data~\cite{mercier2020qfib} & 480,000 & 171,666,931 & iFOD1 & - & - & TCK  & 734.21M  \\
        ISMRM2015~\cite{maierhein2017tractography} & 200,433 & 19,584,878 & synthetic &- & - & TCK & 16.55M  \\
        HCP  (anatomical tracts)~\cite{van2013wu,zhang2018anatomically} & 7,410 & 364,002 & UKF&- & - & TCK & 0.15M \\
        ADHD (whole brain tract)~\cite{zhang2018suprathreshold} & 199,240 & 30,897,382 & UKF &- & - & TCK & 1.23M \\
        dHCP (whole brain tract)~\cite{makropoulos2018developing} & 153,537 & 5,650,084 & UKF &- & - & TCK & 187.08M \\
        \midrule
        HCP~\cite{van2013wu} & 7,410 & 364,002 & UKF& 5 & 5 & VTP & 33.00M \\
        ADHD~\cite{zhang2018suprathreshold} & 19,898,754 & 2,971,986,861 & UKF & 9 & 5& VTP & 149,678.00M \\
        dHCP~\cite{makropoulos2018developing} & 153,537 & 5,650,084 & UKF & 4 & - & VTK & 530.00M\\
        \bottomrule
    \end{tabular}}
\end{table}

We consider the TRAKO, zfib, and qfib data formats for efficient tractography storage. We test these formats with eight different datasets and compute the following metrics to measure compression and data loss. Five datasets only include fiber tracts (Table~\ref{tab:datasets}, top) while three datasets include mapped per-vertex scalars and per-fiber properties (Table~\ref{tab:datasets}, bottom).

Following the qfib paper~\cite{mercier2020qfib}, we use the compression ratio $C_r$. This ratio yields the percentage in reduction of compressed to original size.
\begin{equation} \label{eq:compressionratio}
    C_r = 100 \times (1 - \frac{\textnormal{compressed size}}{\textnormal{original size}})
\end{equation}

Further, to facilitate comparison with other published results, we compute the compression factor $C_f$ to compare the size of original and compressed data.
\begin{equation} \label{eq:compressionfactor}
    C_f = \frac{\textnormal{original size}}{\textnormal{compressed size}}
\end{equation}

TRAKO and qfib do not change the number of points and we calculate individual data loss by measuring point-wise errors as $L^2$-norm. 

\begin{equation} \label{eq:euclidean}
    E = \sum_i|f_i-g_i|,
\end{equation}

for two fiber tracts $f$ and $g$ with the same number of vertices.

We also calculate the endpoint errors by only considering the start and end points of each fiber. This allows to compare with zfib, a method that changes the numbers of fiber points.

\begin{table}[!ht]
    \centering
    \caption{Detailed comparison of qfib (8bit and 16 bit), zfib/Dipy, and TRAKO (JSON and Binary). The first five datasets only contain fiber tracts. TRAKO yields a lower mean error in 4 out of 5 datasets with compression rates of up to 28$\times$. The bottom three datasets include per-vertex scalars and per-fiber properties.}
    \label{tab:results}
    \resizebox{\linewidth}{!}{
    \begin{tabular}{lrrr|llr|llr|lr}
        \toprule
        & \textbf{Size} & \textbf{Ratio} & \textbf{Factor} & \multicolumn{3}{c|}{\textbf{Error}} & \multicolumn{3}{c|}{\textbf{Endpoints Error}} & \multicolumn{2}{c}{\textbf{Timings [m]}} \\
        \textbf{} &  &  $C_r$ & $C_f$~ & ~min & max & mean~ & ~min & max & mean~ & ~enc. & dec. \\
        \midrule
\textbf{qfib-data} & 734.21M\\
~~~qfib (8bit)~\cite{mercier2020qfib} & 22.9M & 96.881\% & 32.064$\times$ & 0.0 & 0.758 & 0.058$\pm$0.023 & 0.0 & 0.74 & 0.038$\pm$0.038 & 476.644 & 65.973\\
~~~qfib (16bit)~\cite{mercier2020qfib} & 44.24M & 93.975\% & 16.597$\times$ & 0.0 & 0.019 & \textbf{0.002$\pm$0.001} & 0.0 & 0.017 & \textbf{0.001$\pm$0.001} & 476.738 & 66.711\\
~~~zfib/Dipy~\cite{presseau2015new} & 118.65M & 83.839\% & 6.188$\times$ & 0.0 & 0.0 & 0.0$\pm$0.0 & 0.0 & 0 & 0.0$\pm$0.000 & 95.14 & 2997.115\\
~~~TRAKO & 46.18M & 93.71\% & 15.899$\times$ & 0.0 & 0.018 & 0.01$\pm$0.003 & 0.0 & 0.018 & 0.01$\pm$0.002 & 273.328 & 190.095\\
~~~TRAKO (Binary) & 34.63M & 95.283\% & 21.199$\times$ & 0.0 & 0.018 & 0.01$\pm$0.003 & 0.0 & 0.018 & 0.01$\pm$0.002 & 272.421 & 188.598\\

\textbf{ISMRM2015} & 16.55M\\
~~~qfib (8bit)~\cite{mercier2020qfib} & 0.98M & 94.103\%& 16.957$\times$& 0.0 & 59.541 & 11.686$\pm$6.327 & 0.0 & 59.522 & 10.501$\pm$10.501 & 269.627 & 45.37\\
~~~qfib (16bit)~\cite{mercier2020qfib} & 1.74M & 89.465\%& 9.492$\times$& 0.0 & 59.316 & 11.61$\pm$6.293 & 0.0 & 59.296 & 10.443$\pm$10.443 & 272.044 & 48.281\\
~~~zfib/Dipy~\cite{presseau2015new} & 8.69M & 47.512\%& 1.905$\times$& 0.0 & 0.0 & 0.0$\pm$0.0 & 0.0 & 0.0 & 0.0$\pm$0.000 & 46.237 & 354.191\\
~~~TRAKO & 1.46M & 91.2\%& 11.364$\times$& 0.0 & 0.233 & \textbf{0.092$\pm$0.027} & 0.001 & 0.229 & \textbf{0.092$\pm$0.015} & 32.803 & 48.85\\
~~~TRAKO (Binary) & 1.09M & 93.401\%& 15.154$\times$& 0.0 & 0.233 & \textbf{0.092$\pm$0.027} & 0.001 & 0.229 & \textbf{0.092$\pm$0.015} & 16.708 & 26.481\\

\textbf{HCP (tracts only)} & 0.15M\\
~~~qfib (8bit)~\cite{mercier2020qfib} & 0.01M & 94.442\%& 17.992$\times$& 0.0 & 18.687 & 0.418$\pm$0.251 & 0.0 & 18.687 & 0.351$\pm$0.351 & 9.432 & 2.847\\
~~~qfib (16bit)~\cite{mercier2020qfib} & 0.01M & 91.362\%& 11.576$\times$& 0.0 & 116.186 & 0.456$\pm$0.321 & 0.0 & 116.186 & 0.451$\pm$0.451 & 9.571 & 3.137\\
~~~zfib/Dipy~\cite{presseau2015new} & 0.08M & 48.524\%& 1.943$\times$& 0.0 & 0.0 & 0.0$\pm$0.0 & 0.0 & 0.0 & 0.0$\pm$0.000 & 1.498 & 0.305\\
~~~TRAKO & 0.01M & 91.385\%& 11.608$\times$& 0.001 & 0.27 & \textbf{0.097$\pm$0.028} & 0.005 & 0.247 & \textbf{0.097$\pm$0.016} & 0.923 & 0.949\\
~~~TRAKO (Binary) & 0.01M & 91.731\%& 12.093$\times$& 0.001 & 0.27 & \textbf{0.097$\pm$0.028} & 0.005 & 0.247 & \textbf{0.097$\pm$0.016} & 1.314 & 1.206\\

\textbf{ADHD Single (tracts only)} & 1.23M\\
~~~qfib (8bit)~\cite{mercier2020qfib} & 0.04M & 96.38\%& 27.624$\times$& 0.0 & 72.832 & 1.762$\pm$1.391 & 0.0 & 71.284 & 1.496$\pm$1.496 & 165.298 & 40.044\\
~~~qfib (16bit)~\cite{mercier2020qfib} & 0.08M & 93.286\%& 14.895$\times$& 0.0 & 120.936 & 4.123$\pm$3.119 & 0.0 & 120.936 & 3.331$\pm$3.331 & 165.486 & 40.681\\
~~~zfib/Dipy~\cite{presseau2015new} & 0.25M & 80.058\%& 5.014$\times$& 0.0 & 0.0 & 0.0$\pm$0.0 & 0.0 & 0.0 & 0.0$\pm$0.000 & 36.811 & 12.235\\
~~~TRAKO & 0.06M & 95.349\%& 21.501$\times$& 0.0 & 0.276 & \textbf{0.08$\pm$0.023} & 0.001 & 0.264 & \textbf{0.079$\pm$0.013} & 61.298 & 40.806\\
~~~TRAKO (Binary) & 0.04M & 96.523\%& 28.76$\times$& 0.0 & 0.276 & \textbf{0.08$\pm$0.023} & 0.001 & 0.264 & \textbf{0.079$\pm$0.013} & 66.261 & 42.501\\

\textbf{dHCP (tracts only)} & 187.08M\\
~~~qfib (8bit)~\cite{mercier2020qfib} & 9.33M & 95.01\%& 20.041$\times$& 0.0 & 53.695 & 0.452$\pm$0.235 & 0.0 & 53.695 & 0.282$\pm$0.282 & 14.954 & 2.027\\
~~~qfib (16bit)~\cite{mercier2020qfib} & 14.68M & 92.154\%& 12.746$\times$& 0.0 & 53.381 & 0.475$\pm$0.375 & 0.0 & 53.381 & 0.442$\pm$0.442 & 15.647 & 2.408\\
~~~zfib/Dipy~\cite{presseau2015new} & 73.68M & 60.616\%& 2.539$\times$& 0.0 & 0.0 & 0.0$\pm$0.000 & 0.0 & 0.0 & 0.0$\pm$0.000 & 23.993 & 2532.927\\
~~~TRAKO & 12.7M & 93.213\%& 14.734$\times$& 0.001 & 0.273 & \textbf{0.152$\pm$0.043} & 0.005 & 0.271 & \textbf{0.152$\pm$0.025} & 9.575 & 5.963\\
~~~TRAKO (Binary) & 9.52M & 94.91\%& 19.645$\times$& 0.001 & 0.273 & \textbf{0.152$\pm$0.043} & 0.005 & 0.271 & \textbf{0.152$\pm$0.025} & 9.091 & 5.921\\

    \bottomrule
    \end{tabular}}
    \resizebox{\linewidth}{!}{
    \begin{tabular}{lrlr}
        \toprule
        &  \textbf{Mean Error} & \textbf{}   & \textbf{Mean Error}\\

        \midrule

        \textbf{HCP}~\cite{van2013wu}, 13.43M, $C_r$: 59.162\%, $C_f$: 2.449$\times$ & \\
        ~~~\textit{Scalars} & & \textit{Properties} \\
        ~~~\begin{tabular}[t]{@{}r@{}}
EstimatedUncertainty ($N$, range:  0.032-15233.791 ) \\
tensor1 ($N\times$9, range:  -0.00095-0.0024 ) \\
tensor2 ($N\times$9, range:  -0.00087-0.0021 ) \\
HemisphereLocataion ($N$, range:  1.0-3.0 ) \\
cluster\_idx ($N$, range:  0-39 ) \\

        \end{tabular} 
        & 
        \begin{tabular}[t]{@{}r@{}}
        
0.135$\pm$0.081 \\
1.121e-07$\pm$2.27e-08 \\
8.73e-08$\pm$1.78e-08 \\
0.0$\pm$0.0 \\
0.246$\pm$0.361 \\

        \end{tabular} 
        & 
        \begin{tabular}[t]{@{}l@{}}
EmbeddingCoordinate ($N$, range:  -4.543-3.047 ) \\
ClusterNumber ($N$, range:  8-665 ) \\
EmbeddingColor ($N$, range:  0-180 ) \\
TotalFiberSimilarity ($N$, range:  199220.9-920767.25 ) \\
MeasuredFiberSimilarity ($N$, range:  0.00179-0.00266 ) \\
        \end{tabular} 
        & 
        \begin{tabular}[t]{@{}r@{}}
0.00026$\pm$3.72188e-05 \\
0.4237$\pm$0.4763 \\
0.8776$\pm$0.4748 \\
8.0194$\pm$4.7547 \\
7.4e-09$\pm$4.5e-09 \\

        \end{tabular}\\

        \textbf{ADHD}~\cite{zhang2018suprathreshold}, 50,462.34M, $C_r$: 66.286\%, $C_f$: 2.966$\times$ & \\
        ~~~\textit{Scalars} & & \textit{Properties} \\
        ~~~\begin{tabular}[t]{@{}r@{}}
            NormalizedSignalEstimationError ($N$, range:  0.0-0.05 ) \\
            EstimatedUncertainty ($N$, range:  0.04-31041.65 ) \\
            RTOP1 ($N$, range:  1.13-23901.94 ) \\
            RTOP2 ($N$, range:  1.32-8651.45 ) \\
            RTAP1 ($N$, range:  -13541.7-7914.96 ) \\
            RTAP2 ($N$, range:  1.11-6820.54 ) \\
            RTPP1 ($N$, range:  0.71-9.88 ) \\
            RTPP2 ($N$, range:  0.71-15.96 ) \\
            SignalMean ($N$, range:  0.0-0.04 ) \\
        \end{tabular} 
        & 
        \begin{tabular}[t]{@{}r@{}}
        
            0.0$\pm$0.0 \\
            0.3$\pm$0.176 \\
            0.04$\pm$0.023 \\
            0.014$\pm$0.008 \\
            0.031$\pm$0.018 \\
            0.01$\pm$0.006 \\
            0.0$\pm$0.0 \\
            0.0$\pm$0.0 \\
            0.0$\pm$0.0 \\
        
        \end{tabular} 
        & 
        \begin{tabular}[t]{@{}l@{}}
            EmbeddingCoordinate ($N\times$10, range:  -3.18-4.93 ) \\
            ClusterNumber ($N$, range:  12-768 ) \\
            EmbeddingColor ($N\times$3, range:  2-180 ) \\
            TotalFiberSimilarity ($N$, range:  149876.58-696306.3 ) \\
            MeasuredFiberSimilarity ($N$, range:  0.0-0.0 ) \\
        \end{tabular} 
        & 
        \begin{tabular}[t]{@{}r@{}}
            0.0$\pm$0.0 \\
            0.0$\pm$0.0 \\
            0.869$\pm$0.511 \\
            5.599$\pm$3.341 \\
            0.0$\pm$0.0 \\
        \end{tabular}\\

        \textbf{dHCP}~\cite{makropoulos2018developing}, 256.31M, $C_r$: 52.799\%, $C_f$: 2.119$\times$ & \\
        ~~~\textit{Scalars} & & \textit{Properties} \\
        ~~~\begin{tabular}[t]{@{}r@{}}
FreeWater ($N$, range:  0.0-1.0 ) \\
tensor1 ($N\times9$, range:  -0.00132-0.0031 ) \\
tensor2 ($N\times9$, range:  -0.00132-0.0043 ) \\
EstimatedUncertainty ($N$, range:  0.0332-196.16 ) \\
        \end{tabular} 
        & 
        \begin{tabular}[t]{@{}r@{}}
1.42e-05$\pm$9.34e-06 \\
2.27e-07$\pm$4.63e-08 \\
2.895e-07$\pm$5.9e-08 \\
0.291$\pm$0.177 \\
        
        \end{tabular} 
        & 
        ~~~~~-
        & \\

        \bottomrule
    \end{tabular}}
\end{table}

\begin{figure}[!hb]
    \centering
    \includegraphics[width=.49\textwidth]{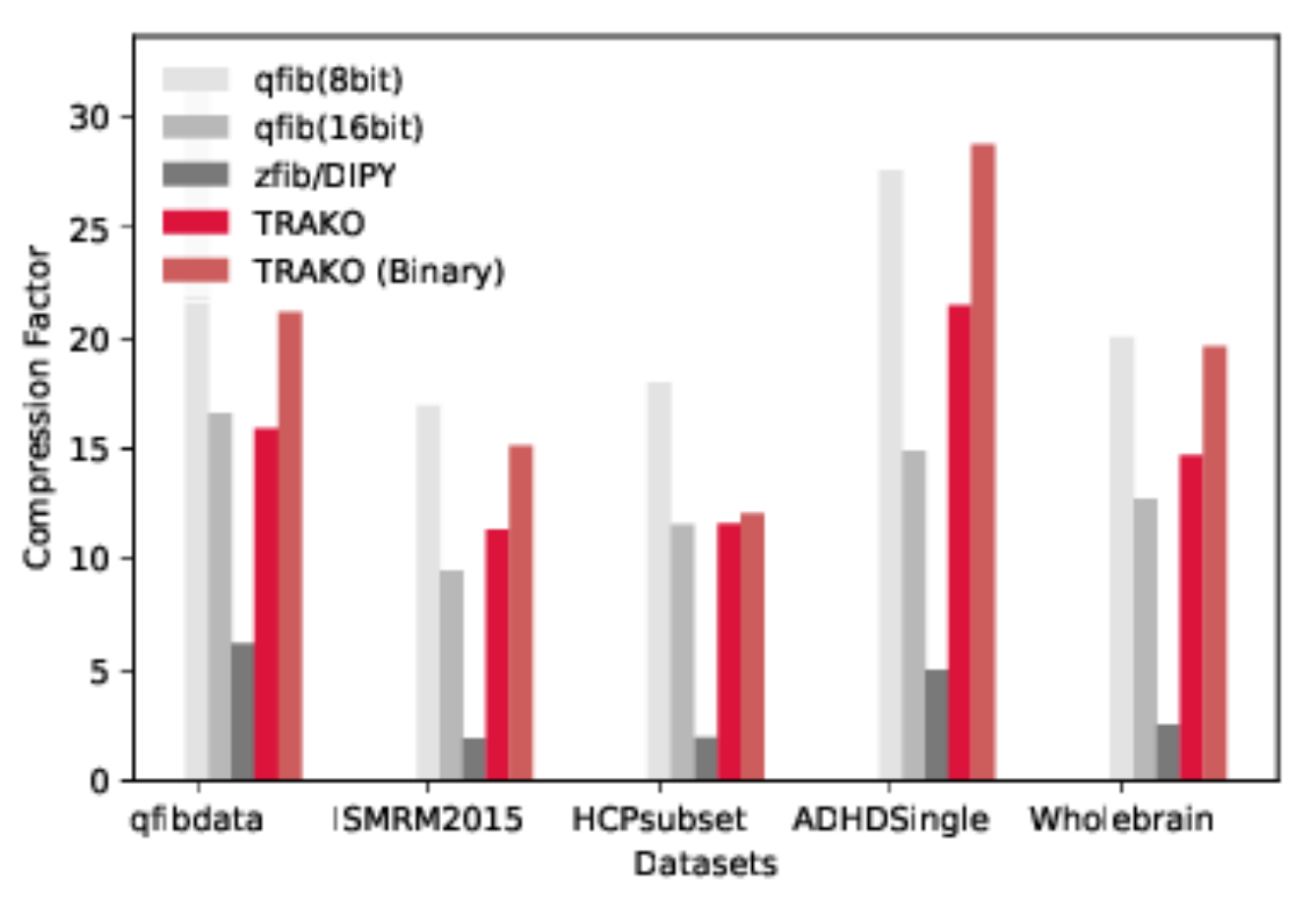}
    \includegraphics[width=.49\textwidth]{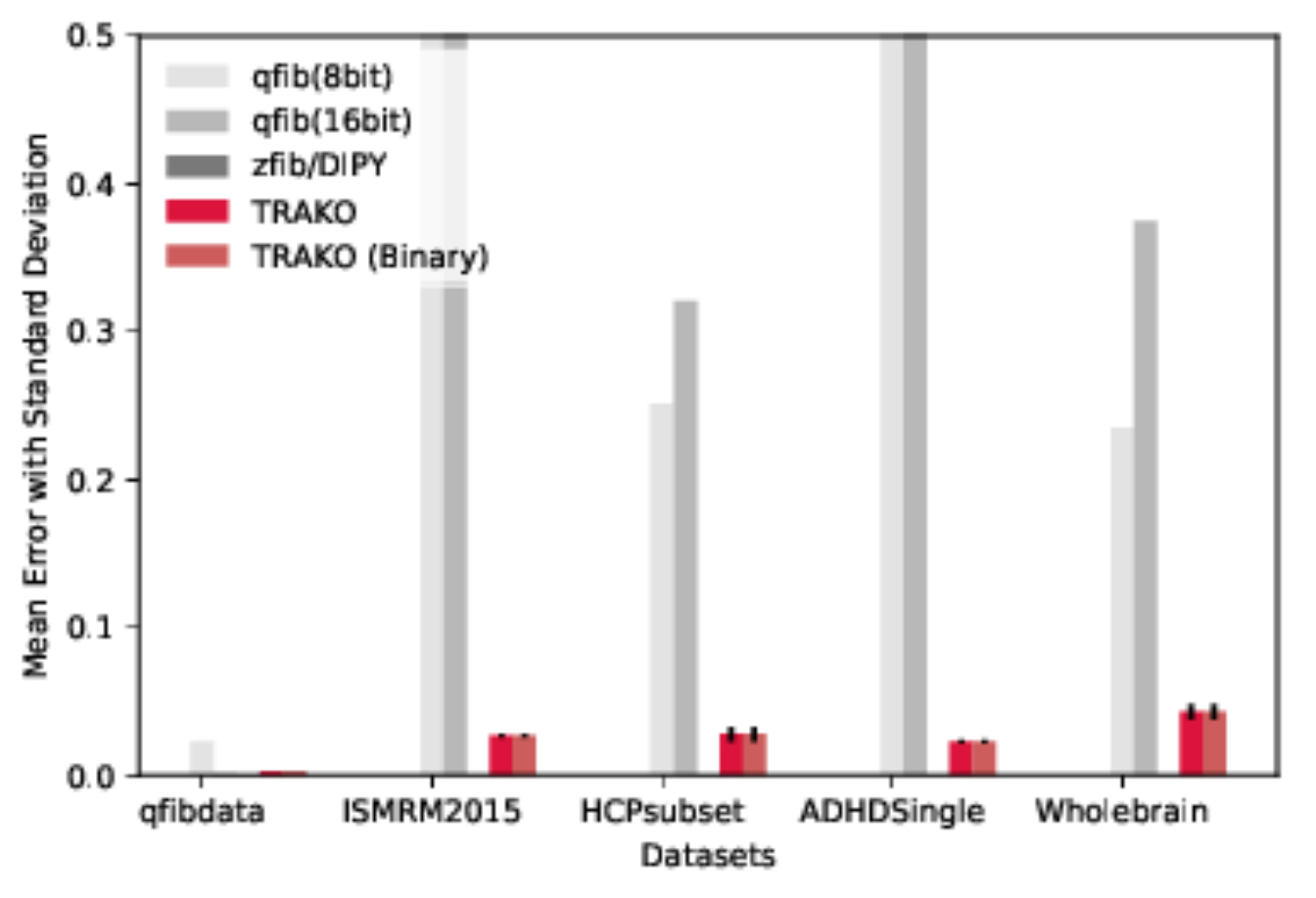}
    \caption{On the five datasets that include only streamlines and vertices, TRAKO produces a comparable compression factor to qfib (and superior to zfib), and in average, a lower mean error (4 out of 5 cases). TRAKO is the only method that supports the three datasets with attached per-vertex scalars and per-fiver properties.}
    \label{fig:performancecomparison}
\end{figure}

\subsection{Sensitivity Analysis}

\subsubsection{Suprathreshold fiber cluster whole brain tractography statistics.}

In this experiment, we assessed if group-wise tractography differences can be preserved using restored data after applying TRAKO (compress and restore). To do so, we performed a suprathreshold fiber cluster (STFC) statistical analysis \cite{zhang2018suprathreshold} on the ADHD dataset to identify group differences in the whole brain tactography between the ADHD and healthy population. The STFC method first performs a data-driven tractography parcellation to obtain white matter fiber parcels (a total of 1416 tract parcels). Diffusion measure of interest, i.e., return-to-the-origin probability (RTOP) \cite{ning2015estimating}, was extracted from each fiber parcel and tested between the two populations using a student t-test. Then, a non-parametric permutation test was performed to correct for multiple comparisons across all fiber parcels. Overall, the output of the analysis includes STFCs, i.e. a fiber cluster of multiple fiber parcels that are significantly different when comparing the RTOP diffusion measure (p $<$ 0.05).

We performed the STFC analysis on the original tractography data, as well as the restored data. Each individual fiber parcel was compressed and decompressed using TRAKO using the default options, yielding the compression factors and error rates as reported in Table~\ref{tab:results}. In the original data, there were two sets of STFCs (corrected p values 0.015 and 0.035, respectively). In the restored data, the same sets of STFCs were identified (corrected p values 0.009 and 0.028, respectively), suggesting good performance of TRAKO on preserving group-wise tractography differences. 

\subsubsection{Bhattacharyya overlap distance.}
To ensure TRAKO does not alter the fiber tract points, we have additionally implemented the Bhattacharyya analysis and computed the overlap score ($B$) to quantify the agreement between the original and restored tract points \cite{Rathi2013,CetinKarayumak2018}:

$B = \frac{1}{3}\left (  \int \sqrt{P_o(x)P_r(x)}dx + \int \sqrt{P_o(y)P_r(y)}dy +  \int \sqrt{P_o(z)P_r(z)}dz \right )$,
with the ground truth probability distribution $P_o(.)$ of the original fiber tract, $P_r(.)$ the probability distribution from the restored fiber tract, and the fiber coordinates $\mathbf{x} = (x,y,z)\in\mathbb{R}^3$. $B$ becomes 1 for a perfect match between two fiber bundles from original and restored data and 0 for no overlap at all. 

We performed the Bhattacharyya overlap distance analysis on the corpus callosum (CC) tract, which was parcellated using \cite{zhang2018suprathreshold} for both original and restored fiber tracts. We then computed the overlap score between the original and restored CC in all subjects (0.99$\pm$1.6231e-04). The very high overlap between original and restored tract points indicates that TRAKO can successfully preserve this information during compression.

\section{Conclusions}

We have introduced TRAKO, a data format for tractography fiber tracts that allows for high data size reduction with low information loss. Built-off the glTF community standard to allow immediate GPU processing, TRAKO is also the only data format that compresses tractography data with attached per-vertex scalars and per-fiber properties. In the future we plan to use TRAKO to distribute tractography datasets, thus reducing download times for interactive visualization and data transmission costs for large-scale analysis. To encourage community adoption, we release TRAKO and our results as free and open research at \texttt{https://github.com/haehn/trako/}.

\clearpage
\printbibliography

\end{document}